\documentstyle{mn}

\title{A radio-jet -- galaxy interaction in 3C441}
\author[Lacy et al.\ ]
       {Mark Lacy, Steve Rawlings, Katherine M.\ Blundell \& 
Susan E.\ Ridgway \\
        Astrophysics, Department of Physics, Keble Road, Oxford, OX1 3RH}
\date{}

\pubyear{1997}

\begin{document}

\maketitle
\begin{abstract}
Multi-wavelength imaging and spectroscopy of the $z=0.708$ radio galaxy
3C441 and a red aligned optical/infrared component are used to show that the 
most striking aspect of the radio-optical ``alignment effect'' in this
object is due to the interaction of the radio jet with a companion galaxy
in the same group or cluster. The 
stellar population of the red aligned continuum component
is predominately old, but with a small post-starburst
population superposed, and it is surrounded by a low
surface-brightness halo, possibly a
face-on spiral disc. The [O{\sc iii}]500.7/[O{\sc ii}]372.7 emission line
ratio changes dramatically from one side of the component to the other, with 
the low-ionisation material apparently having passed through the bow shock of
the radio source and been compressed. A simple model for the interaction
is used to explain the velocity
shifts in the emission line gas, and to predict that the ISM of the interacting
galaxy is likely to escape once the radio source bow shock has passed though.
We also discuss another, much fainter, aligned component, 
and the sub-arcsecond scale
alignment of the radio source host galaxy. 
Finally we comment on the implications of our explanation
of 3C441 for theories of the alignment effect.
\end{abstract}


\section{introduction}
The consequences of radio jets impacting on density inhomogeneities
have been invoked to explain many properties 
of high redshift radio sources, such as bending and asymmetries in arm length.
Evidence for this is seen in the form of  
correlations between emission-line gas and the side of the radio 
galaxy with the shorter arm-length, or higher depolarisation
(McCarthy, van Breugel \& Kapahi 1991; Liu \& Pooley 1991). 
The extent to which this is related to the so-called ``alignment effect''
whereby the axis of extended optical continuum and line emission is 
found to be co-aligned with the radio axis, is unclear, although
many attempts to explain radio-optical alignments involve
a dense external medium. Such a medium is needed, for example, to act
as the source of a scattering surface
for quasar light from the AGN (Bremer, Fabian \& Crawford 1997), 
or as a medium for jet-induced star formation (e.g.\ Best, Longair \& 
R\"{o}ttgering 1996 \& references therein).

3C441 is a $z=0.708$ radio galaxy with an asymmetric radio
structure, which appears to be a rare 
example of a radio source with a red aligned component outside the radio 
lobes. Such a component is clearly hard to obtain in either jet-induced 
star formation or scattered quasar models.
This red component is seen just beyond the end of the shorter, north-west 
radio lobe [component `c' of Eisenhardt \& Chokshi (1990); see Fig.\ 1]. 
This lobe appears to possess a 
radio jet which bends round to the west at the tip of the lobe. 
Just to the south of the red component, mostly within the radio lobe, is a 
arc-shaped clump of emission line gas seen in the [O{\sc ii}]372.7 emission 
line image of McCarthy, van Breugel \& Spinrad (1994). Spectroscopy of the 
[O{\sc ii}] emission line
by McCarthy, Baum \& Spinrad (1996) shows that this emission line 
material is redshifted by $\approx 800\;$km s$^{-1}$ with respect to the 
radio galaxy, and weak emission extends to just beyond the red component. 

In this paper, we use our own spectroscopy, infrared and radio
imaging, and archive data from the {\em Hubble Space Telescope} {\em
(HST)} (also presented in Best, Longair \& R\"{o}ttgering 1997c) to 
attempt to explain the properties of 3C441 in terms of models for the 
interaction of radio jets with their environments. In particular, we address 
the problems of the origin of the red continuum light from the aligned 
component and the properties of the extended emission line region.

We assume an Einstein -- de Sitter cosmology with a Hubble constant
$H_0=50\;$km s$^{-1}$Mpc$^{-1}$.

\section{Observations}
We obtained an optical spectrum with ISIS on the 4.2-m William Herschel
Telescope (WHT) on 1993 August 20
with the aim of establishing the nature of component `c'. Both the 
red and blue arms of ISIS were employed, the red arm detector being the
EEV5 CCD and the blue arm detector the TEK1. Observations were taken 
using a 3-arcsec wide slit at PA 45 deg.\ aligned so as to pass
through both the host galaxy (`a' on Fig.\ 1) 
and the knot `c'. The airmass was between 1.02 and 1.06 and 
the seeing 0.9 arcsec
during the observations, which consisted of a total of 3600s exposure 
in the blue arm and 3500s exposure in the red arm.

Observations in $J$- and $K$-bands were made with IRCAM3 on
the United Kingdom Infrared Telescope (UKIRT) on 1996 November 6 for 
20 min and 40 min respectively. Conditions
were non-photometric. The seeing was 0.8 arcsec with the fast guider employed.
The $K$-band image showed pattern noise in the form of stripes across
the image, these were removed with the {\sc iraf} task {fit1d}.

A $J$-band spectrum was obtained with CGS4 on UKIRT on 1994 June 17
covering the region of the H$\alpha$ line. Conditions were non-photometric,
with some light cirrus present. 3C441 was observed for 3200s with the
long camera and the 75 l mm$^{-1}$ grating in second order. The 2-pixel
(3-arcsec) wide slit was aligned at the same PA as for the optical 
observations. The combined optical and infrared  spectra are shown in 
Fig.\ 2. 

Radio observations were made in B-array with the Very Large Array radio 
telescope at 8.3 GHz on 1992 January 11. The exposure time was 1950s and the 
bandwidth 50 MHz. The data were calibrated for phase and amplitude 
in the standard manner and analysed in {\sc aips}. One of the two IFs
was removed due to interference problems. A correction for Ricean
bias was made to the polarisation map using the {\sc aips} task {\sc polco}.

{\em HST} 
imaging data were obtained from the archive, and consisted of a total of 
1700s exposure in the 555W and 785LP filters with the WFPC2. The radio galaxy 
was centered on the WF3 CCD, on which the scale is 0.1 arcsec pixel$^{-1}$. 
The filters both contain strong redshifted emission lines: the F555W 
filter contains the [O{\sc ii}]372.7 emission and the F785LP filter
the [O{\sc iii}]495.9 and 500.7 lines. 
Fig.\ 1 shows an overlay of the 8 GHz radio map contours on greyscales of the
two {\em HST} images.

\setlength{\unitlength}{1pt}
\begin{figure*}
\begin{picture}(600,300)
\put(-180,-285){\includegraphics{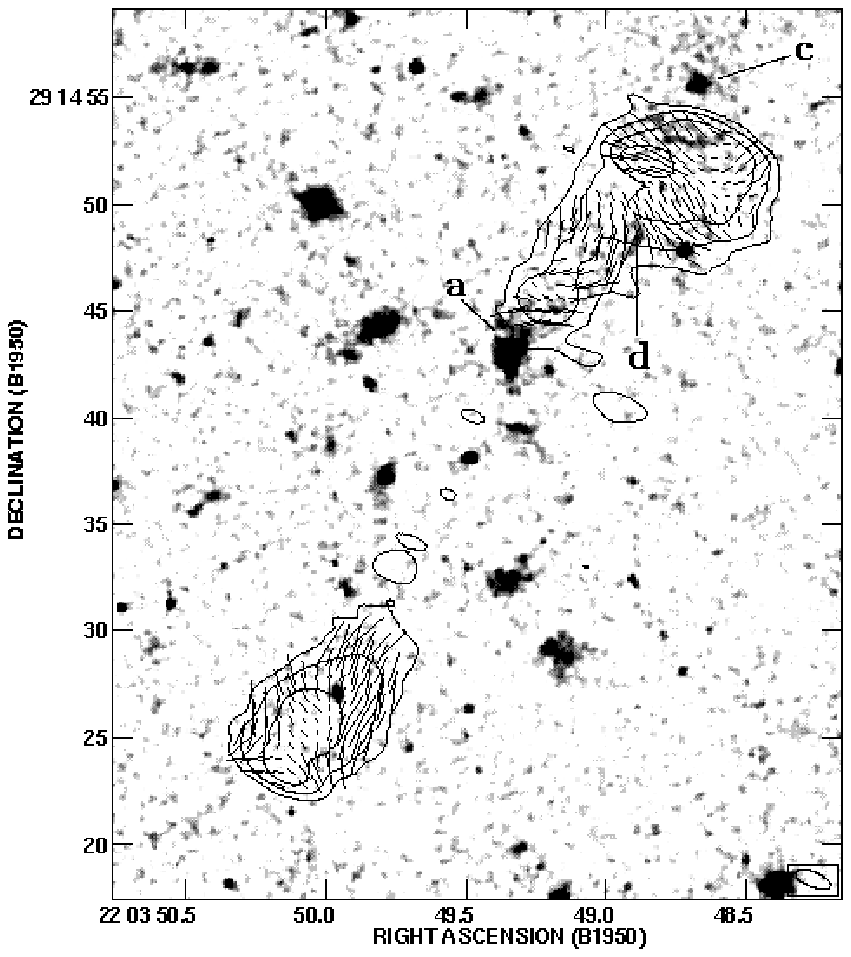}}
\put(90,-285){\includegraphics{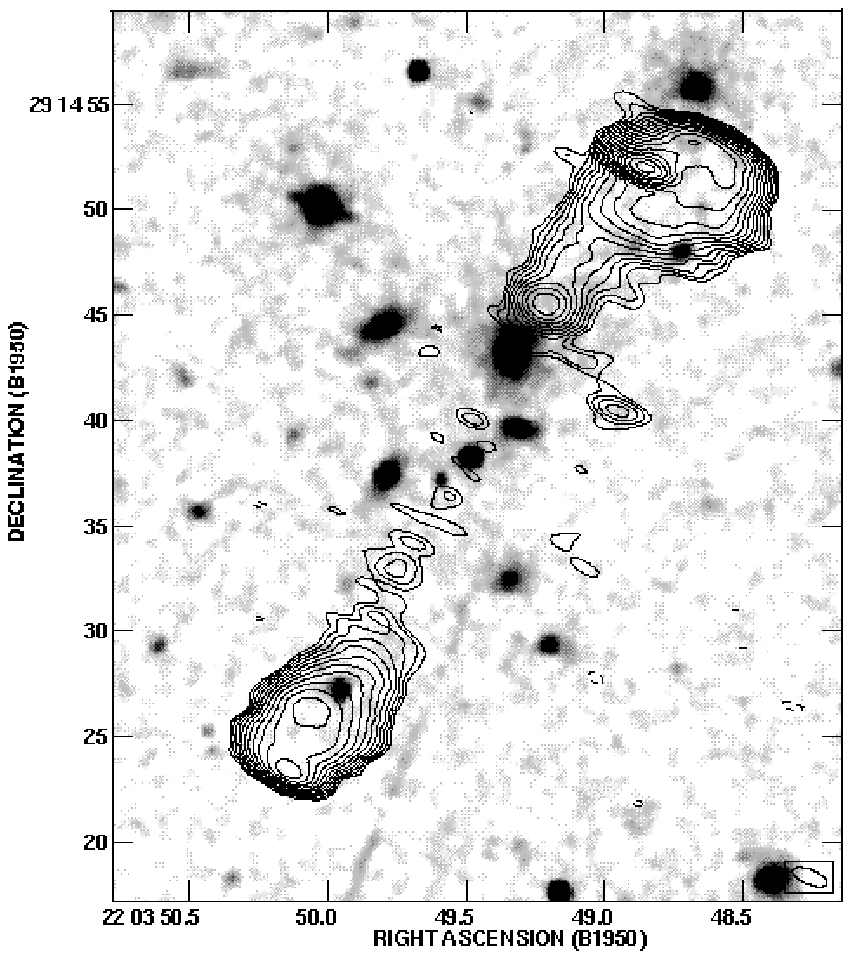}}
\put(0,290){\bf (a)}
\put(270,290){\bf (b)}
\end{picture}
\caption{
Overlay of the WFPC2 images of 3C441 on our VLA 8 GHz B-array radio map. (a)
The F555W image smoothed with a $\sigma=0.1$ arcsec gaussian, 
with radio contours and magnetic field polarisation vectors
superposed. The optical components referred to in the text are labelled.
Contours are spaced at intervals separated by factors of four from 0.4
mJy beam$^{-1}$, and
the polarisation vectors are scaled so that a length of 1 arcsec represents
25 per cent polarisation. (b) The F785LP image smoothed with 
a gaussian with $\sigma=0.2$ arcsec, with radio 
contours again superposed. Contours are
spaced at intervals separated by factors of $2^{1/2}$ starting at 0.3
mJy beam$^{-1}$. The radio beam FWHM is indicated in the bottom right
corner of both plots.}
\end{figure*}

\begin{table}
\caption{Emission line fluxes from 3C441}
\begin{tabular}{lrlrc}
Emission line & Flux in `a'& $z_{\rm a}$ & Flux in `c'& $z_{\rm c}$ \\ 
              &/$10^{-19}$& &/$10^{-19}$ & \\
              &${\rm Wm^{-2}}$     & &${\rm Wm^{-2}}$             & \\
\quad [Ne{\sc iv}]242.4& 1.3 & 0.702 & $<$0.5 & - \\
\quad [Ne{\sc v}]342.6 & 1.7 & 0.7095 & $<$0.5 & - \\
\quad [O{\sc ii}]372.7  & 4.5 & 0.7094 &4.5&0.7132\\
\quad [Ne{\sc iii}]396.8& 1.7 & 0.7084 & $<$0.5 & - \\
\quad H$\gamma$           & 0.6 & 0.7088 & $<$0.5 & - \\
\quad H$\beta$            & 1.4 & 0.7083 & $<$0.9 & - \\
\quad [O{\sc iii}]495.9 & 7.6 & 0.7084 & $<$0.9& - \\
\quad [O{\sc iii}]500.7 & 21.9& 0.7084 &2.7&0.7136\\
\quad H$\alpha$+[N{\sc ii}]&17  & 0.709  & $<$10  &    \\
\end{tabular}

Notes: $z_{\rm a}$ and $z_{\rm c}$ are the mean redshift of the emission
line in components `a' and `c' respectively.
No aperture corrections have been applied. The 
optical continuum flux matched the broad-band flux measurement in
a 3-arcsec diameter aperture to
20 per cent, but the infrared continuum flux was low by 0.5 mag,
suggesting that an aperture correction of $\approx 1.6$ should be
applied to H$\alpha$+[N{\sc ii}] flux. No attempt was made to deblend
the H$\alpha$+[N{\sc ii}] complex due to the low signal-to-noise of
the detection.
\end{table}


\begin{figure*}

\begin{picture}(400,405)
\put(-60,-140){\includegraphics{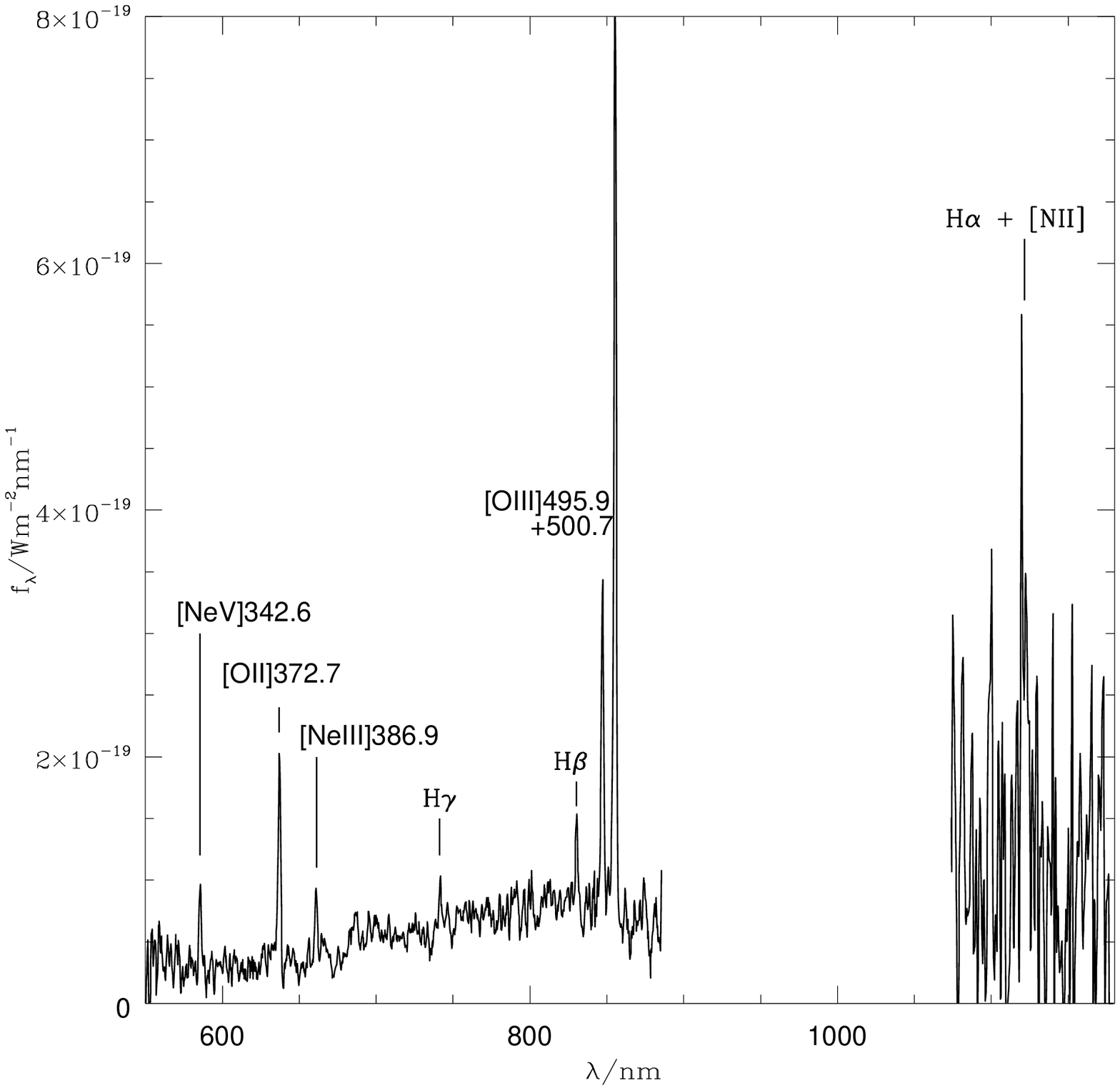}}
\end{picture}

\caption{The spectrum of the radio galaxy `a'. The ISIS 
red-arm optical spectrum has
been smoothed with a 5-pixel (1.4 nm) box-car filter and the CGS4 near-infrared
spectrum by a 9-pixel (7.2 nm) one.}
\end{figure*}

\section{Discussion}

\setlength{\unitlength}{1mm}
\begin{figure}
\begin{picture}(75,75)
\put(0,-20){\includegraphics{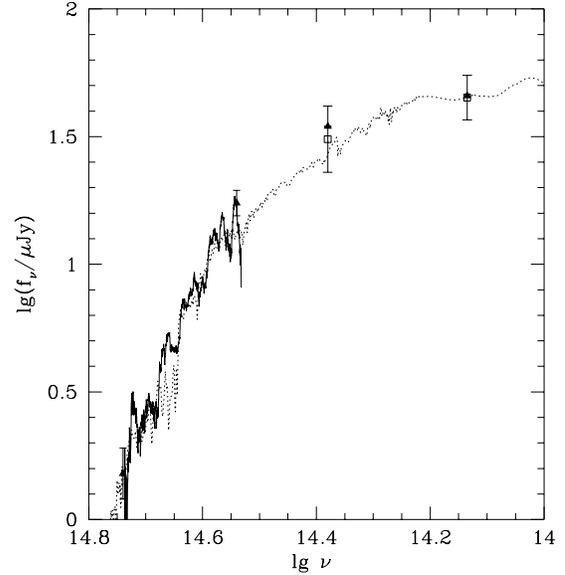}}
\end{picture}
\caption{\small{The SED of the compact red knot `c'. 
The continuous line is from our own spectroscopy, the 
near infrared points from Eisenhardt \& Chokshi (1990) (open squares) and 
our own near-infrared and optical photometry (solid triangles; note the $J$ and
$K$ points have no associated error bars as the data were non-photometric).
The dotted line is the model spectral energy distribution 
of a 6-Gyr old galaxy in which star formation 
proceeded at a constant rate in a 1-Gyr burst before ceasing
(Bruzual \& Charlot 1993).}}
\end{figure}

\subsection{The continuum knot}

Component `c' in Fig.\ 1a and in Eisenhardt \& Chokshi (1990), 
the red aligned knot beyond the northwest radio hotspot,
is slightly resolved on the {\em HST} images.  
Its spectral energy distribution (SED) seems to be 
consistent with that of an old stellar population (Fig.\ 3), although  
the amplitude of the 4000\AA$\;$break seems lower than that of 
the model 6-Gyr old
stellar population (corresponding approximately to the age of the Universe  
in our assumed cosmology) which is otherwise a good fit to the SED.

A stellar origin for the light is indeed confirmed by a close inspection
of the spectrum of `c' (Fig.\ 4) in which stellar absorption features,
in particular the ``G-band'' from CH absorption in cool stellar envelopes
can be seen at $z=0.714$, redshifted by $1000\, {\rm km\, s^{-1}}$ 
with respect to the radio galaxy ($z=0.708$). 
Also visible are H$\delta$ and the calcium H and K lines, 
the latter possibly blended with H$\epsilon$. The weakness of the 
4000\AA$\;$break, despite the apparent dominance of the SED by an old stellar 
population may then be explained by the existence of a small 
post-starburst population of A-stars, and indeed the 
detection of H$\delta$ absorption in our spectrum is consistent with this.
Such populations have been detected 
in other AGN host galaxies and companions (Tadhunter, Dickson \& Shaw 1996; 
Canalizo \& Stockton 1997), and may trace mergers or interactions in small 
clusters or groups (Zabludoff et al.\ 1996). 

Whether there is a link between 
a starburst $\sim 10^8$yr ago in a close companion or the host itself
and the triggering of the AGN is unclear -- certainly if so there must be
a delay between the starburst and the development of a radio source if 
typical radio source lifetimes are $\sim 10^7$yr (Alexander \& Leahy 1987).
In the case of 3C441, component `c' is $\approx 125$ kpc from the host galaxy,
so it is conceivable that if its relative velocity with respect to the
host in the plane of the
sky is of the same order as its relative velocity along the line of sight, it
could have been close
enough to interact $10^8$yr ago. There is, however, no sign of a corresponding
starburst in the host galaxy, and E+A galaxies are relatively common in 
high-$z$ clusters (Dressler \& Gunn 1990; Caldwell \& Rose 1997). Furthermore,
the high velocity encounter implied by this may not have have produced 
sufficient disruption to initiate the starburst.

\setlength{\unitlength}{1mm}
\begin{figure}
\begin{picture}(75,75)
\put(0,-20){\includegraphics{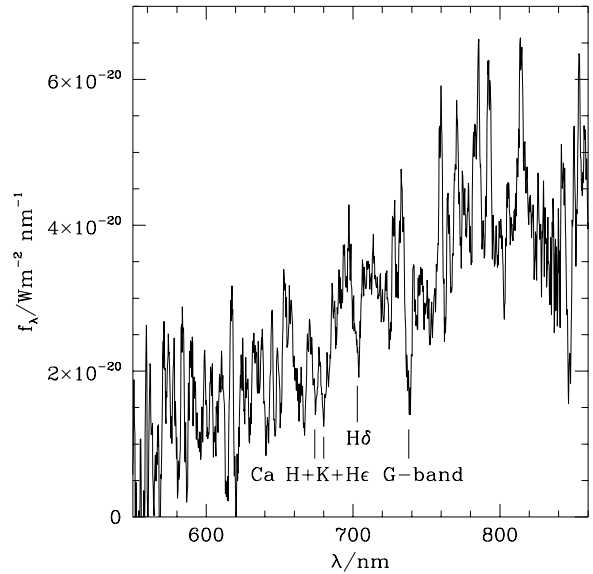}}
\end{picture}
\caption{\small{The red-arm spectrum of knot `c' of 3C441, showing the
stellar absorption features mentioned in the text. The raw spectrum has
been smoothed by a 9-pixel (2.4 nm) box-car filter and corrected for 
atmospheric absorption.}}
\end{figure}

\subsection{The emission line gas in `c'}

Contour plots of the 2-D spectra around the [O{\sc ii}]372.7 and 
[O{\sc iii}]500.7 emission lines are shown in Fig.\ 5. There is a 
striking change in the ionisation and luminosity across the position of the  
continuum knot: the side nearest to the radio galaxy has a high luminosity,
and a low ionisation as measured by the ratio of 
[O{\sc iii}]500.7/[O{\sc ii}]372.7 emission lines ($0.35 \pm 0.1$). 
In contrast, beyond the 
knot the [O{\sc iii}]500.7/[O{\sc ii}]372.7 ratio is much higher ($>10\pm 3$) 
and the total luminosity lower by a factor of $\approx 6$. 
There is a velocity gradient of $\approx 300$ km s$^{-1}$ in the 
line emission across the continuum knot, in the sense that the side nearest 
the radio galaxy is blueshifted, and that
on the far side is, within errors of $\approx 100\, {\rm km\,
s^{-1}}$, at rest 
with respect to the
starlight in `c'. There is also a faint tail of yet more 
highly blueshifted emission (up to $\approx 600\, {\rm km\, s^{-1}}$)
on the side nearest the radio galaxy, pointing towards the radio galaxy. 

Note that there is no sign of emission lines in Fig.\ 4, which used a narrow
extraction about the continuum peak. In contrast, in Fig.\ 5 the 
extended emission-line flux from `c' is clearly visible and is quite
strong when integrated over the entire emission region (Table 1).

\setlength{\unitlength}{1mm}
\begin{figure*}
\begin{picture}(150,60)
\put(-25,-200){\includegraphics{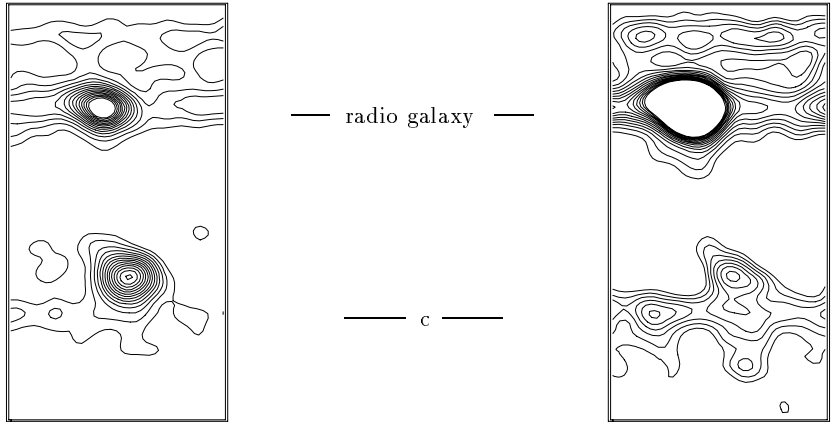}}
\end{picture}
\caption{\small{2-D spectrum of 3C441 with the slit aligned along the 
axis joining the radio galaxy and `c'. Left: the [O{\sc ii}] emission line
with wavelength on the horizontal axis and distance along the slit 
vertically. The radio galaxy is to the top and component `c' below it. Right: 
the [O{\sc iii}]5007 emission line. The figures are  
$11.2\, {\rm nm} \times 30\, {\rm arcsec}$ in size (11.2 nm $\approx$ 
5280 km s$^{-1}$ close to [O{\sc ii}] and $\approx$ 3930 km s$^{-1}$ close to 
[O{\sc iii}]). The contour levels are spaced at intervals of $2\times 10^{-21}
{\rm Wm^{2}nm^{-1}}$ per pixel, starting from $2\times 10^{-21}
{\rm Wm^{2}nm^{-1}}$, each pixel was $0.28 {\rm nm} \times 0.33 {\rm arcsec}$
in size.}}
\end{figure*}

\subsection{The radio structure}

The asymmetric radio structure of 3C441 is interesting in the context
of the models to explain the aligned emission. The asymmetry in jet
brightness either side of the nucleus is very pronounced, and can be
interpreted either as an asymmetry produced by Doppler boosting, or in
terms of 3C441 having a radio structure transitional between FRI and
FRII, with one side FRII-like and the other more FRI-like. 
The lack of a radio central component, commonly seen in radio
galaxies with Doppler boosted one-sided jets (e.g. 3C22, Rawlings et
al.\ 1994), argues strongly in favour of the latter explanation, and
indeed the flaring of the jet just to the NW of the host galaxy is 
reminiscent of the ``Mach disk'' structure seen in the M87 jet (Owen, Hardee
\& Cornwell 1989). 

\subsection{The radio galaxy}

The spectrum of `a', the host galaxy of the radio source is shown in
Fig.\ 2. It is a typical moderately-high ionisation narrow-line 
radio galaxy spectrum, with strong emission lines superposed on a
stellar spectrum dominated by an old stellar population with a strong 
4000\AA$\;$break. 

Close inspection of the images shows, however, that 
even in this case, where the integrated light is dominated by old
stellar populations there is evidence for morphological peculiarity. 
In particular there is a distinct blue aligned component to the
south east of the host galaxy peak (Fig.\ 6). Whether this
represents a spiral arm, a merger remnant or some form of radio 
source-induced aligned component is unclear. As its separation
from the radio galaxy is only about 0.5-arcsec, it is hard 
to tell from the spectra whether it is line or continuum 
dominated, but the more diffuse material extending $\approx 2$-arcsec
to the south of the radio galaxy is definitely continuum dominated. 
Although much weaker
relative to the smooth underlying host galaxy in the F785LP image, the
aligned component
is nevertheless visible, along with some more diffuse aligned emission
on the other side of the peak, to the northwest. 

In Fig.\ 7, the position angle of the host galaxy (derived from the
second moments of the flux distribution) is plotted for various
isophotes and apertures. This shows two interesting aspects of the
alignment. First, the alignment persists into
the $K$-band, suggesting the aligned light is fairly red. Second,
in the {\em HST} images, there is evidence for 
isophotal twisting as the aperture size is increased to include the
inner aligned components highlighted in Fig.\ 6 (PA $\sim 150$ deg), and
later the low surface brightness emission round the host at PA $\sim
0$, seen best in Fig.\ 1. The low-surface brightness material to the
northwest may be 
mostly line emission though, as the emission-line image of McCarthy et al.\
(1995) shows that the [O{\sc ii}] emission is also roughly aligned
along PA 0.

\setlength{\unitlength}{1pt}
\begin{figure}
\begin{picture}(200,100)
\put(-240,-350)
{\includegraphics{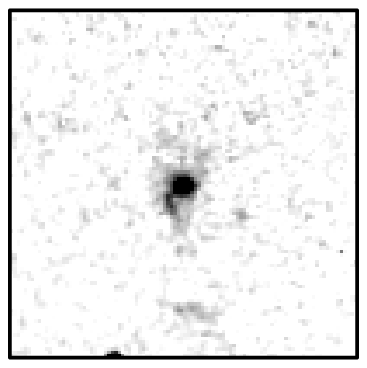}} 
\put(-120,-350){\includegraphics{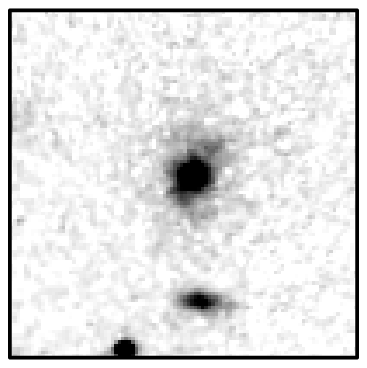}}
\put(0,80){(a)}
\put(120,82){(b)}
\end{picture}
\caption{Close-up of the host galaxy (`a') greyscaled so as to show
the aligned components near the nucleus: (a) F555W image; (b) F785LP
image. Both images are 10 arcsec square and have been smoothed with a 
$\sigma = 0.05$ arcsec gaussian.}
\end{figure}

\setlength{\unitlength}{1pt}
\begin{figure*}
\begin{picture}(500,200)
\put(-165,-250){\includegraphics{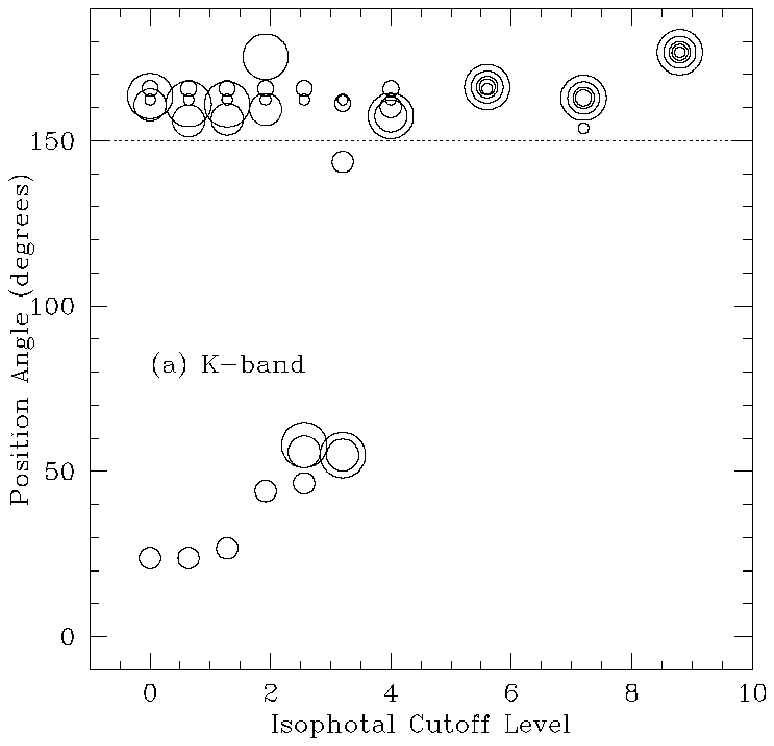}}
\put(15,-250){\includegraphics{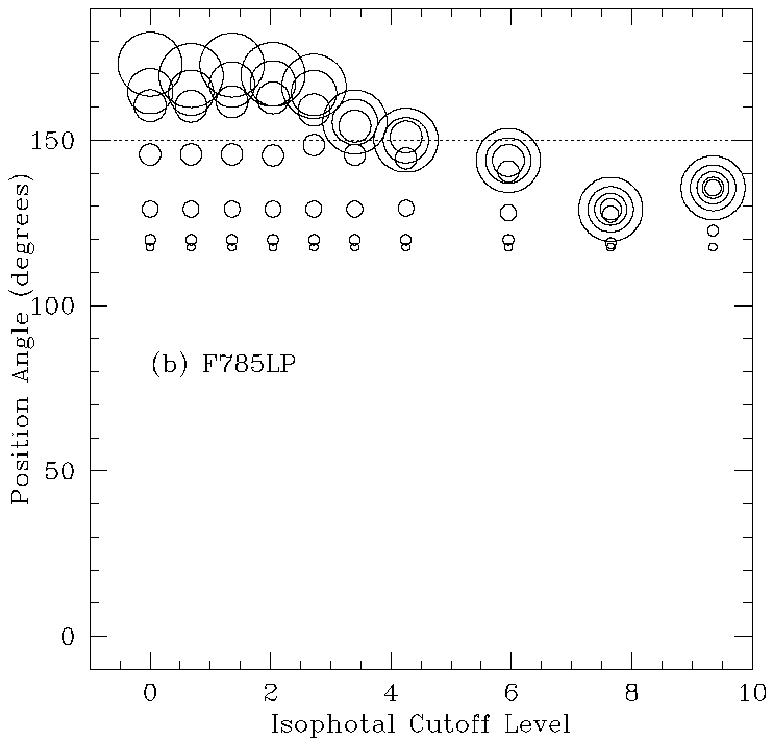}}
\put(195,-250){\includegraphics{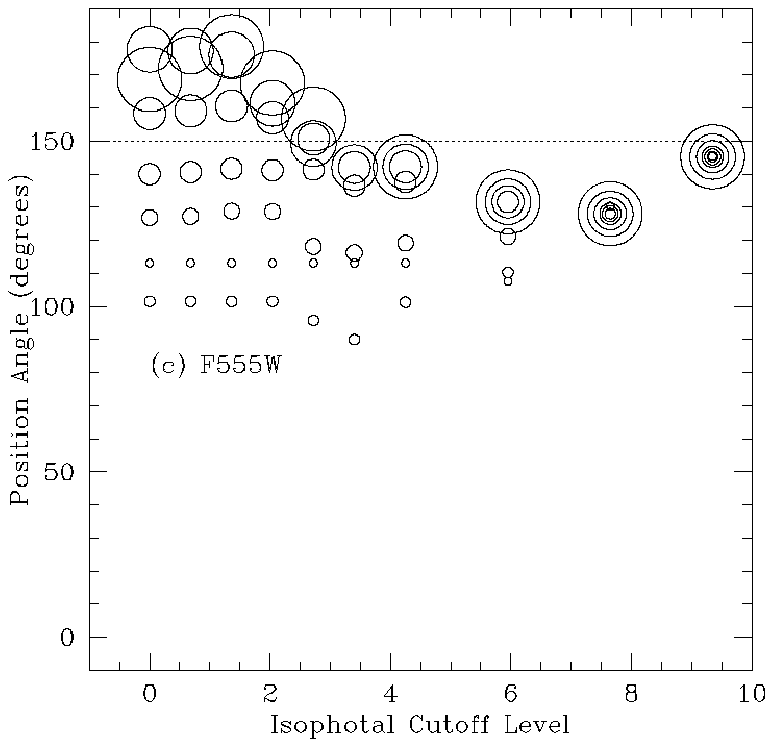}}
\end{picture}
\caption{Position angle as a function of aperture size and isophotal
cutoff for the host galaxy of 3C441. The isophotal cutoff
level is in units of the sky noise, and the aperture radii (indicated by
circles of increasing size) are 0.6, 0.9, 1.1, 1.7 and  2.4
arcsec in (a), and 0.3, 0.4, 0.6, 0.8, 1.2, 1.7, and 2.4 arcsec in (b) and
(c). The radio source PA (defined as the PA of the line joining the
hotspots) is indicated by a dotted line.}
\end{figure*}

\subsection{The relationship of the radio and optical components}

The relative astrometry was performed
using the results of the APM scans of the Palomar Sky Survey plates. 
The positions of four stars were used to align the 
corners of the radio map with the optical images. The uncertainty in the 
overlay from the scatter in the fit was $\approx 0.3$ arcsec. 
This astrometry places the host galaxy (`a' in Fig.\ 1a) just to the SE of the
first appearance of the northern radio jet. Component `d' is apparently
aligned along the 
jet direction, although positioned to the side of it. The point of 
deflection of the jet is approximately coincident with the peak of the
[O{\sc ii}]372.7 emission, about 3-arcsec SE of the peak of the continuum 
knot `c'. In the F555W image (Fig.\ 1a), there is an arc of emission just to 
the north of the north-west radio hotspot and apparently centered on `c'. 
Fig.\ 1b shows the F785LP image; here there is a halo of diffuse emission  
around `c'. 

By subtracting a spectrum centered on the continuum knot in an aperture 
1.7-arcsec wide from the total spectrum of the aligned component `c' (in 
a 6.8-arcsec wide aperture), we have been able to estimate the line 
contribution to the continuum magnitudes measured for the nebular region 
surrounding `c'. In both filters this is $\approx 10$ per 
cent overall, but in the regions of brightest line emission in the interaction
region to the south of `c' our spectrum
suggests that the line contribution of [O{\sc ii}] to the F555W flux
rises to dominate the overall flux in the filter, consistent with the arc of 
emission seen in the F555W region just above the 
radio hotspot consisting entirely of line emission. The continuum
emission present in this extraction is bluer than the emission from
the knot.

The linear object `d' (Fig.\ 8) is reminiscent
of the aligned component in 3C34 (Best, Longair \& R\"{o}ttgering 1997a). 
Like the object in 
3C34, it has no emission lines visible either in our spectra or in the 
narrow-band image of McCarthy et al.\ (1994), but is well aligned with
the radio structure and lies within the radio lobe. Its optical--near infrared 
colours are bluer than `c' ($K=20.8; J>22.7; m_{785}=23.3; m_{555}=25.1$ in 
3-arcsec diameter apertures, where
$m_{785}$ and $m_{555}$ are AB magnitudes in the two {\em HST} images), 
but show a break between the 785LP and 555W 
filters, consistent with a 4000$\;$\AA$\;$break at the redshift of the
radio source.

\setlength{\unitlength}{1pt}
\begin{figure}
\begin{picture}(200,100)
\put(-548,-740){\includegraphics{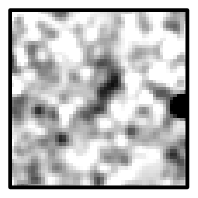}} 
\put(-430,-740){\includegraphics{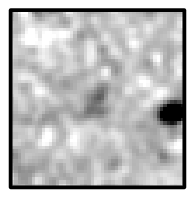}}
\put(0,80){(a)}
\put(120,80){(b)}
\end{picture}
\caption{Close-up of component `d': (a) F555W image; (b) F785LP
image. Both images are 5 arcsec square and have been smoothed with a 
$\sigma = 0.1$ arcsec gaussian.}
\end{figure}

\subsection{Other companion objects}

There are several objects around the radio galaxy which appear to 
be members of a group or cluster around it. 3C441 is in a sample of radio
galaxies whose clustering properties we are currently evaluating, and a 
formal estimate of the clustering amplitude, $B_{\rm gq}$, 
for this radio source 
will be presented in Wold et al.\ (in preparation). For the purposes
of this paper, we simply compared the counts of galaxies with magnitudes
$m_1$ to $m_1+3$ (where $m_1$ is the magnitude of the radio source
host galaxy) in the object frame of the F785LP image (the WF3 CCD) with 
the average of those in the two side frames (WF2 and WF4). 
This revealed an excess of $11.5 \pm 7.1$ galaxies, indicating the 
possible presence of a cluster, but not at a high confidence level. Clearly 
though this is likely to be an underestimate of the cluster richness
as many cluster members may be outside the restricted field of the CCD, and 
present on the other frames, increasing the estimate of the background count.

\section{Interpretation of the aligned structures}

\subsection{Component `c'}

We argue here that the evidence presented above allows us to deduce that
3C441 includes an example of a radio-jet -- galaxy interaction. 
[Best et al.\ (1997c) have also proposed a jet--galaxy interaction at
`c', but were unable to confirm it using the {\em HST} imaging data alone.]

First, the evidence that the northern radio jet is being deflected by an 
increase in density just south of knot `c' is given by the fact that 
the radio jet appears to bend round to the west $\approx 3$ arcsec SE
of `c', consistent with simulations of radio jets interacting
with gas clouds (Higgins, O'Brien \& Dunlop 1997). The polarisation B-field
vectors (Fig.\ 1a) are also consistent with an interaction scenario, tending
to be perpendicular to the emission-line arc. The relatively high polarisation
of the interaction region (up to 30 per cent) suggests that we are seeing the
interaction through the radio lobe (contrast PKS 2250-41; Clark et al.\ 1997,
in which the polarisation of the interacting lobe at 8 GHz is only 2.6 per 
cent, so in that case the emission line gas is almost certainly
covering the radio lobe). Provided the rotation measure is 
$\leq 200$ rad m$^{-2}$
in the rest-frame of the source, the angle of polarisation will be rotated by 
$\leq 5$ deg.

Second, the evidence that the radio jet is shocking the emission-line
gas to the south of knot `c' is strong. The fact that emission-line
gas is present beyond the interaction region shows that
photoionisation is probably
responsible for most, if not all, of the line emission. The main 
photoionisation mechanism could either be photoionsation by the
nucleus, or photoionisation by hot, post-shocked gas. Hot stars in 
H{\sc ii} regions of the galaxy could also contribute to the
photoionising flux. Irrespective of the ionisation mechanism,  
compression of the shocked gas would lead to a decrease in the
ionisation parameter (Clark et al.\ 1997), and hence the difference in 
ionisation between the two sides of knot `c'.
The velocity structure of the emission lines is also consistent with 
shocking, with the emission being blueshifted on the radio galaxy side of the 
continuum knot, suggesting that the radio
jet is approaching the observer. The mean velocity of the knot, a redshift 
of $+1000\, {\rm km\, s^{-1}}$ is assumed to  
come from the random motion of the galaxy in a cluster around
3C441. 

Third, the halo-like appearance of the diffuse emission around a central 
continuum knot and the distribution of [O{\sc iii}]500.7 emission in the
spectrum is hard to explain unless the stars and gas are associated with 
a galaxy. This galaxy has an red old or E+A-type central concentration of 
stars with a surrounding lower surface-brightness disc or halo, in which there 
is a significant quantity of gas. The spectrum of the diffuse emission 
shows that the SED of the extended continuum is
consistent with that of a late-type spiral, with rest-frame $U-B \approx -0.2$
(Coleman, Wu \& Weedman 1980). The most straightforward interpretation of 
this is therefore that we are seeing the interaction of the radio jet
with a face-on
spiral galaxy in the same cluster, with the bulge component providing the 
central high surface brightness knot of red light (with little gas to provide
line emission from this region) and the gas and stars in the 
surrounding disk providing the diffuse halo of emission. 
The [O{\sc ii}] emission, however, appears to trace the bow-shock, 
presumably due to 
compression behind the shock increasing the density and lowering the 
ionisation parameter. The size of the disc (44 kpc in our assumed 
cosmology) is typical of the size of a spiral disc seen in H{\sc i}
(e.g.\ Broeils \& van Woerden 1994). The problem of survival of the gas halo 
or disc in the cluster environment that may surround 3C441 suggests that the
galaxy may only recently have fallen into the cluster.

\subsection{A simple model for the interaction}

We consider a simple model 
in which the radio jet interacts with a two-phase interstellar 
medium of the spiral galaxy. Similar models have been developed to 
investigate the interactions of supernova shocks with the interstellar 
medium, and the case of the interaction of radio jets with a two-phase
medium has been discussed by Rees (1989). These models show that the 
shock wave propagates though the hot phase, initially by-passing the cold 
clouds (in which the shock speed is much lower). The velocity given     
to the cold clouds can be estimated by considering the effect of the 
pressure differential as the shock passes. For a cloud of radius $r$
and a jump in pressure $\Delta P$ across a shock of speed $v_{\rm sh}$
the velocity given to the cloud,
\[ v_{\rm cl} \sim  r^2 \Delta P \left( \frac{r}{v_{\rm sh}} \right)
\frac{1}{\rho_{\rm cl} r^3} \sim \frac{\Delta P}{\rho_{\rm cl} M c_{\rm h}} \]
where $\rho_{\rm cl}$ is the density of the cold clouds, $M$ the 
Mach number of the shock (in the pre-shock medium) and 
$c_{\rm h}$ the speed of sound in the hot-phase gas -- i.e.\ independent 
of the size of the cloud. From the limiting 
forms of the Rankine-Hugoniot relations in the case of a strong 
shock, 
\[ \Delta P \sim P_1 M^{2} \]
where $P_1$ is the preshock gas pressure. Hence  
\begin{eqnarray*}
 v_{\rm cl} & \sim & \frac{P_1 M}{\rho_{\rm cl} c_{\rm h}} \sim 
\frac{c_{\rm cl}^2}{c_{\rm h}^2} v_{\rm sh}\\
            & \sim & 100 \frac{T_{\rm cl}/10^4{\rm K}}{T_{\rm h}
            /10^6{\rm K}} (v_{\rm sh}/10^4{\rm km\, s^{-1}})\, {\rm
km\, s^{-1}}
\end{eqnarray*}
where $T_{\rm cl}$ and $T_{\rm h}$ are the temperatures of the cold and 
hot phases respectively. If we take $v_{\rm sh}=v_{\rm rs}\sim 0.03c$, 
typical for a radio source advance speed (Alexander \& Leahy 1987), this 
process can easily account for the 
modest velocity shifts seen in the gas to the south of the knot relative
to the unshocked gas in the north.

The ionisation parameter (here defined as the dimensionless ratio of the
number density of photoionising photons to the total number density of
hydrogen atoms, $U=n_{\rm ph}/n_{\rm H}$) will change across the shock as
$n_{\rm H}$ is increased ($n_{\rm ph}$ is likely to be similar either side
of the knot whether the ionisation is due to shocks or to the nucleus). 
Typically [O{\sc iii}]500.7/[O{\sc ii}]372.7 is
expected to be a linear function of $U$ over the range we are considering 
(e.g.\ Penston et al.\ 1990). To obtain the observed change in the 
[O{\sc iii}]500.7/[O{\sc ii}]372.7 ratio therefore requires an enhancement
of $n_{\rm H}$ in the postshock region by a factor of $\sim 30$. This 
demonstrates the importance of the two-phase assumption: if only one
phase were present the Rankine-Hugoniot relations predict a density 
enhancement of at most a factor $\approx 4$. The cloud gas should also be 
heated by the compression following the shock, but deeper spectroscopy is 
required to measure the [O{\sc iii}]436.3 temperature diagnostic line either 
side of the shock.

Numerical simulations (e.g.\ Stone \& Norman 1992, who examine the 
physically-similar situation of supernova shocks) show that in fact the 
hot phase, which is moving at a high speed with respect to the cold 
clouds interpenetrates the cold medium and eventually 
(within a few sound-crossing times of the cloud) causes turbulent
mixing of the hot and cold phases. This is probably required to produce the 
high velocity tail of blueshifted emission line gas. 
Gas with high velocity shifts 
and widths ($\stackrel{>}{_{\sim}} 1000 {\rm km\, s^{-1}}$) is seen in
a number of high-$z$ radio galaxies, e.g.\ 3C368, 
and probably arises from this mechanism (Stockton, Ridgway \& Kellogg
1996; Tadhunter 1997).

\subsection{Component `d'}

The apparent close correlation of the position angle of `d'
with the direction of the radio structure may of course be
coincidental, and `d' may well be a companion, foreground or background
galaxy which just happens to align with the radio axis. In section 3.5 
we showed that the colours of `d' were at least consistent with it being in the
same group or cluster as the radio galaxy and `c', but there is no other 
evidence to associated `d' with these two galaxies. Nevertheless, we can 
speculate on what `d' might be if its alignment is not a coincidence. 

One possible explanation is that `d' may be a tidal remnant of an encounter 
between the radio galaxy and `c' (see Section 3.1). If, however, as
previously argued any interaction of the radio galaxy and `c' is unlikely
to have been very disruptive, this leaves the possibility that the 
alignment of `d' is caused by a process involving the radio source, most 
plausibly either starlight produced by jet-induced star formation, or optical
synchrotron emission. 

If `d' is synchrotron emission,  
it is not emission from the jet centre, which clearly passes to one side. 
Emission from a boundary layer of the jet cannot be ruled out,
however, and indeed an interaction of the boundary layer of the jet
with a galaxy or gas clump might be
expected to be a region of high magnetic field and current particle 
acceleration, in a manner analogous to that seen in supernova remnants
(e.g.\ Anderson et al.\ 1994). We nevertheless 
cannot rule out a jet-induced star formation origin for this component, 
on the lines of
that proposed by Best et al.\ (1997a) 
for the similar component in 3C34. Probably the best argument against this
(apart from the lack of emission lines) is 
that the object is parallel to the current jet direction, not the line
joining it to the host galaxy as one might naively expect if the galaxy 
was formed as a result of an interaction with the head of the expanding
radio source.

\subsection{The host galaxy}

Although relatively subtle, as Fig.\ 7 demonstrates, the light from
the host of 3C441 is aligned along the direction of the radio jet in
both {\em HST} passbands (which straddle the 4000\AA$\;$break) and in the 
near-infrared. The fraction of aligned light (defined here as the fractional
excess of emission within $\pm 22.5$ deg.\ of the radio axis) is
largest in the F555W image, at about 20 per cent, falling to ten per
cent in the F785LP image and only four per cent in the $K$-band. Thus although
3C441 is one of the least ``active'' [by the definition
of Best et al.\ (1997b)] galaxies in 3C, and has no detectable
polarisation in the infrared to a limit of $\approx 5$ per cent
(Leyshon \& Eales 1997), we cannot 
eliminate the possibility that the aligned light is scattered. 
Nebular continuum is unlikely to contribute though, because
the observed limit on the H$\alpha$ emission suggests
that nebular continuum accounts for $<1$ per cent of the total $K$-band flux,
and, although in the optical emission lines may contribute to the 
aligned components, there is no strong emission line in $K$-band.

The aligned structures themselves are not closely associated with
visible radio emission, rendering both synchrotron and inverse-Compton
mechanisms (Daly 1992) unlikely. Jet-induced star formation, however, 
cannot be ruled out
here, although the colour difference between the aligned emission
north and south of the galaxy argues somewhat against this, as in the 
jet-induced star formation model stellar populations at equal
distances from the nucleus should have similar colours. However, differential
dust extinction either side of the nucleus could be responsible for the 
colour difference.

The aligned structure seen particularly in the F555W image 
(Fig.\ 6a) is a particularly intriguing feature of this
object, reminiscent of the more spectacular features in
3C280 (Ridgway \& Stockton 1997). 
As it seems fairly blue, it may represent an AGN-produced 
component, or pre-existing material brightened by the interaction with
the radio jet, for example nebular continuum emission from shocked
gas, and it may therefore contribute little to the near-infrared
aligned light. 

The presence of a good near-infrared alignment may  
imply a link between the position angle of the underlying old
stellar population and the radio jet axis, perhaps produced by a
selection effect (Eales 1992). Evidence in favour of this is the 
surprisingly good correlation often seen in 3C radio galaxies 
between the infrared PA and the radio axis. Dunlop \& Peacock (1993) 
even claim that it is better than that seen in the optical. Although 
subsequent {\em HST} studies have shown some of Dunlop \& Peacock's  
infrared alignments to be fortuitous (e.g.\ 3C175.1; Ridgway \& 
Stockton 1997), it is also clear that in some cases
(e.g.\ 3C280; Ridgway \& Stockton 1997), the infrared emission may be well
aligned even after subtraction of some extended aligned components seen
in the optical. 

Alignments between cD galaxies and the projected
distribution of galaxies within clusters have been noted by Rhee, van 
Haarlem \& Katgert (1992). Given the common occurence of clusters
around radio sources at $z\sim 0.5$, it seems not unlikely that the
central galaxy is aligned with the cluster potential, and the
selection effect pointed out by Eales (1992) will then tend to favour aligned
radio sources [but see West (1994) 
for an alternative explanation of
the alignment of the radio source with the cluster/cD axis].

\section{Implications of radio-source -- galaxy interactions}

3C441 is a striking example of the ``alignment effect'' whereby the 
radio and optical axes of $z>0.7$ powerful radio galaxies are frequently
found to be co-aligned. In this case, if we are correct, the most 
spectacular aspect of the alignment, that of knot `c' with the radio axis,
has a relatively simple explanation --- namely the interaction of the 
radio jet with another galaxy in the same group or cluster. A jet-galaxy
interaction has also been invoked to explain the properties of the aligned
emission in the $z=0.3$ radio galaxy 
PKS 2250-47 (Clark et al.\ 1997). How relevant 
these are to the more problematic cases seen at higher redshifts (and  
usually higher radio powers too) is unclear. 
For example, the  
host galaxy and knot `c' of 3C441 are $<5$ per cent polarised
(Tadhunter et al.\ 1992; Leyshon \& Eales 1997), but in many examples
of high-$z$ radio galaxies the aligned light has a polarisation of 
$\approx 10$ per cent. Also, there are very few examples 
where the aligned component is outside the radio lobes. 

Jet-induced star formation, the process whereby radio jets impacting
on gas clouds promote star formation, appears {\em not} to be happening
in the aligned component `c' of 3C441. Although the appropriate conditions 
appear to be in place (namely a powerful radio jet impacting on dense clouds 
of gas), there is apparently no significant extra 
blue continuum associated with the interaction region 
that might come from a population of very young stars
(the youngest stellar population detectable in `c' is the spiral disc
population, which is equally bright on both sides of the central
knot). However, 
it may simply be that the interaction has occurred too recently for the 
stars to begin to form. 

The effects of the interaction on galaxy `c' may be dramatic.
The post-shock temperature of the hot ($10^6$K) component of the ISM
of `c' will be 
raised as it passes through the shock by a factor $\sim M^2/4$. Assuming 
$v_{\rm sh}\sim 10^4 {\rm km\, s^{-1}}$, 
this is a factor of $\sim 10^3$. Mixing
with the cold medium and radiative losses will cool it somewhat, but it is 
likely that the resulting gas will still have thermal velocities exceeding 
the galaxy escape velocity (corresponding to $T\sim 10^7$K), and will 
ultimately evaporate into the intracluster medium (ICM). Thus radio source 
interactions may provide a method of stripping gas from cluster galaxies,
and thereby adding metal-enriched gas to the ICM.

The importance of jet -- companion galaxy interactions in producing
radio -- optical alignments is still to be fully quantified. It is
likely that interactions of this sort produce large amounts of line
emission and some nebular continuum, and, if they occur close to the
nucleus, the continuum emission will be further brightened by
scattering of the hidden quasar light. Any old stellar population in
the companion galaxy contributes light in the near infrared, and, if
the selection effect pointed out by  Eales (1992) is effective, 
we would expect to
see more companion galaxies along the radio axes than elsewhere. 
Furthermore, if
radio galaxies form by the accumulation of smaller sub-units as in the
``bottom-up'' structure formation hierarchy, close companions
will be a common feature of high-$z$ radio galaxy hosts.

\section*{Acknowledgements}
We thank Chris Simpson for assistance with the UKIRT observations.
We are also grateful to the support staff at the WHT and UKIRT for their
help with the observations. We also thank the referee for a number 
of useful comments. This paper was partly based on observations made 
with the NASA/ESA Hubble Space Telescope, obtained from the data archive at 
the Space Telescope Science Institute. STScI is operated by the Association of
Universities for Research in Astronomy, Inc. under the NASA contract 
NAS 5-26555. The National Radio Astronomy Observatory
is a facility of the National Science Foundation  operated under 
cooperative agreement by Associated Universities Inc.
The WHT is operated on the island of La Palma by the Royal 
Greenwich Observatory in the Spanish Observatorio del Roque de los Muchachos
of the Instituto de Astrofisica de Canarias, and the UKIRT by the Royal 
Observatory Edinburgh on behalf of the UK PPARC. We are grateful to Mike Irwin
and Richard McMahon for making the APM scans of the Palomar Sky Survey 
generally available.

\end{document}